\documentclass[reprint,showpacs,amsmath,amssymb,aip,apl]{revtex4-1}

\usepackage{graphicx}
\usepackage{amssymb}
\usepackage{dcolumn}
\usepackage{textcomp}
\usepackage{amsmath}

\begin{document}

\title[Probing the TLS Density of States]{Probing the TLS Density of States in SiO Films using Superconducting Lumped Element Resonators}

\author{S. T. Skacel}
\affiliation{Physikalisches Institut,
Karlsruher Institut f\"ur Technologie, Wolfgang-Gaede-Stra{\ss}e 1, D-76131
Karlsruhe, Germany}
\affiliation{Institut f{\"u}r Mikro- und Nanoelektronische Systeme,
Karlsruher Institut f\"ur Technologie, Hertzstra{\ss}e 16, D-76187
Karlsruhe, Germany}

\author{Ch.~Kaiser}
\affiliation{Institut f{\"u}r Mikro- und Nanoelektronische Systeme,
Karlsruher Institut f\"ur Technologie, Hertzstra{\ss}e 16, D-76187
Karlsruhe, Germany}

\author{S.~Wuensch}
\affiliation{Institut f{\"u}r Mikro- und Nanoelektronische Systeme,
Karlsruher Institut f\"ur Technologie, Hertzstra{\ss}e 16, D-76187
Karlsruhe, Germany}

\author{H.~Rotzinger}
\affiliation{Physikalisches Institut,
Karlsruher Institut f\"ur Technologie, Wolfgang-Gaede-Stra{\ss}e 1, D-76131
Karlsruhe, Germany}

\author{A.~Lukashenko}
\affiliation{Physikalisches Institut,
Karlsruher Institut f\"ur Technologie, Wolfgang-Gaede-Stra{\ss}e 1, D-76131
Karlsruhe, Germany}

\author{M.~Jerger}
\affiliation{Physikalisches Institut,
Karlsruher Institut f\"ur Technologie, Wolfgang-Gaede-Stra{\ss}e 1, D-76131
Karlsruhe, Germany}

\author{G.~Weiss}
\affiliation{Physikalisches Institut,
Karlsruher Institut f\"ur Technologie, Wolfgang-Gaede-Stra{\ss}e 1, D-76131
Karlsruhe, Germany}

\author{M.~Siegel}
\affiliation{Institut f{\"u}r Mikro- und Nanoelektronische Systeme,
Karlsruher Institut f\"ur Technologie, Hertzstra{\ss}e 16, D-76187
Karlsruhe, Germany}

\author{A.~V.~Ustinov}
\affiliation{Physikalisches Institut,
Karlsruher Institut f\"ur Technologie, Wolfgang-Gaede-Stra{\ss}e 1, D-76131
Karlsruhe, Germany}
\affiliation{Russian Quantum Center, 100 Novaya St., Skolkovo, Moscow region, 143025, Russia}

\date{\today}

\begin{abstract}
We have investigated dielectric losses in amorphous SiO thin films under operating conditions of superconducting qubits (mK temperatures and low microwave powers). For this purpose, we have developed a broadband measurement setup employing multiplexed lumped element resonators using a broadband power combiner and a low-noise amplifier. The measured temperature and power dependences of the dielectric losses are in good agreement with those predicted for atomic two-level tunneling systems (TLS). By measuring the losses at different frequencies, we found that the TLS density of states is energy dependent. This had not been seen previously in loss measurements. These results contribute to a better understanding of decoherence effects in superconducting qubits and suggest a possibility to minimize TLS-related decoherence by reducing the qubit operation frequency. 
\end{abstract}




\maketitle


For the realization of a quantum information processor based on superconducting circuits, physical mechanisms responsible for decoherence have to be cleared up \cite{Schoen_review}. Microwave resonators are key ingredients of superconducting quantum circuits. Dielectric thin films are widely used in qubit fabrication as insulating layers and overlay capacitors of the resonators. Lumped capacitors are often employed in qubits to shunt Josephson junctions in order to adjust their operation frequency. It is well established \cite{Martinis} that two-level tunneling systems (TLS) in insulating films contribute to dielectric losses at microwave frequencies and thus often appear as dominant decoherence sources, in particular for superconducting phase qubits. In this paper, we experimentally study microwave properties of amorphous dielectric thin films embedded in superconducting lumped element resonators.

Dielectric losses induced by TLS in the qubit working regime have been extensively studied in recent years and good agreement with the predictions of the TLS tunneling model has been found \cite{Martinis, Lindstroem, Macha, Gao2, Barends1}. However, most of the investigations were done using coplanar waveguide resonators, where TLS and thus losses originate from surfaces \cite{Lindstroem, Macha, Gao2, Barends1}. The bulk losses in dielectric films may also play a significant role for a variety of qubit designs and till now remain less understood. Furthermore, there is little known about the energy dependence of the TLS density of states $n(E)$ in typical qubit materials. A better understanding of this matter would allow to improve qubit coherence and lead to a better qubit design and operation. In the standard tunneling model (TM) of glasses, a dependence of $n(E)$ on energy is not excluded \cite{Phillips1}, but characteristic glassy behavior refers to an essentially constant density of states \cite{Phillips2, Anderson}. So far, a slight dependence of $n(E)$ on energy had been extracted from specific heat measurements \cite{Stephens}, but could be explained in terms of the TM taking into account the time scales of the measurements \cite{Phillips1}. Earlier investigations of the sound velocity in glassy thin a-SiO$_{x}$ films in the kHz \cite{Gaganidze} and 300~MHz \cite{Haumeder} ranges validate the constancy of the density of states $n(E)$ predicted by the TM.
 
In this work, we investigate the dielectric response to applied electric fields at frequencies of a few GHz. Previous measurements of dielectric losses in this frequency range using coplanar waveguide resonators  showed no energy dependence of $n(E)$ \cite{Lindstroem}. In coplanar resonators, only a fraction of the electric field is stored in the TLS hosting material under investigation, so that filling factors $F$ have to be estimated \cite{Lindstroem, Macha, Gao2, Barends1}. Uncertainties in estimating this filling factor are rather large and vary from resonator to resonator, thus preventing a detailed study of the energy dependence of n(E) in those measurements. Our goal here is to investigate the energy dependence of the TLS density of states in the volume of typical amorphous insulating thin films used for qubit fabrication. As dielectric material, we have chosen a-SiO, which we have previously used for fabrication of phase qubits \cite{Kaiser2}. In order to obtain direct comparison of losses at different frequencies, we developed a broadband measurement setup employing multiplexed lumped element resonators, a power divider and a broadband cryogenic amplifier.

The TM for amorphous solids was developed \cite{Phillips2, Anderson} to explain the low-temperature thermal properties of glasses that are drastically different from crystals \cite{Zeller}. TLS with electric dipole moments dominate the low-temperature dielectric properties. They give, in particular, rise to resonant coupling to electric microwave fields at the frequency $f=E/h$ \cite{Schickfus, Phillips1}. Resonant absorption of microwaves depends on the occupation number difference of the two states with energy splitting $E$. Thus, it will be effective (i) for temperatures $T\lesssim E/k_{\rm B}$ and (ii) for microwave powers $P$ small enough to not affect the occupation numbers noticeably. At high temperatures or microwave power levels beyond a critical value $P_{\rm c}$, the resonant absorption vanishes. 

The dielectric loss in the resonant absorption regime ($E=hf$) is given by \cite{Enss}   
\begin{eqnarray}
  \label{eqn:tandeltastrong}
	\tan\delta &=& \tan\delta_{\rm sat}+\alpha\cdot\mathcal F(P)\cdot\tanh\left(\frac{E}{2 k_{\rm B}T}\right), \\
 	\label{eqn:fP}
 	\mathcal F(P)&=&\left(1+\frac{P}{P_{\rm c}}\right)^{-\frac{1}{2}}, \\
  \label{eqn:alpha}
	\alpha &=&\frac{\pi\cdot n(E)\cdot p^{2}}{3\varepsilon_{0}\varepsilon_{\rm r}}.
\end{eqnarray}

\begin{figure}
\centering
\includegraphics[width=0.85\linewidth]{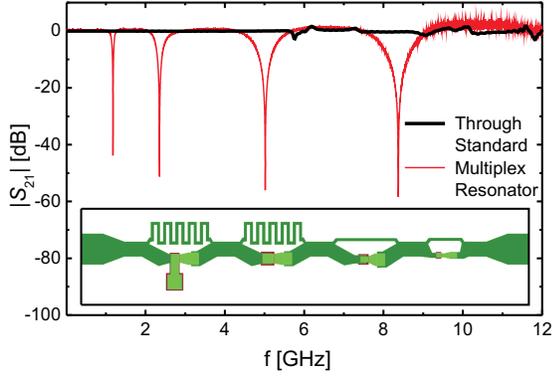}
\caption{Four SiO resonators multiplexed in series (thin line), covering a frequency range from 1.1 GHz to 8.2 GHz. The 8.2 GHz resonator could not be measured at low temperature because it was damaged when it was mounted on the cooling stage. Cold calibration of the setup was possible using a broadband power combiner (thick line). Inset: Schematic design of the multiplexed lumped element resonator.}
\label{fig:8ressetup}
\end{figure}

\noindent
Here, $\tan\delta_{\rm sat}$ represents the loss level at saturation, where both TLS levels are equally occupied. For glasses in general, $\tan\delta_{\rm sat}$ is dependent on frequency and temperature. $\varepsilon_{0}$ and $\varepsilon_{\rm r}$ are the vacuum permittivity and the relative permittivity of the investigated material, respectively. $p$ denotes the mean TLS dipole moment while the factor 1/3 accounts for the three spacial dimensions. Clearly, $p$ may depend on microscopic details of the individual TLS. Since dielectric measurements, however, integrate over a large number of TLSs, in the TM only an average value for $p$ is retained, being a constant for a given substance. Accordingly, we assume the mean dipole moment $p$ to be constant in the investigated frequency range between 1 GHz and 5 GHz. 
Thus, the analysis with Eqs.~(1)-(3) of dielectric loss measurements at various frequencies allows to extract the energy dependence of the density of states of TLS, $n(E)$.\\
\begin{figure}
\includegraphics[width=0.67\linewidth]{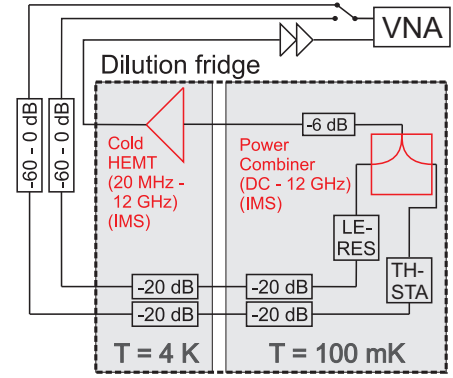}
\caption{Schematics of the broadband loss measurement setup with the possibility of cold calibration. Two identical microwave feed lines are attenuated at different temperatures. Measured signals pass a broadband power combiner and are amplified at 4.2~K as well as at room temperature. One branch contains the multiplexed lumped element resonators (LE-RES) while the other one contains a through standard (TH-STA). Cold amplifier and power combiner are designed and built at the Institute of Micro- und Nanoelectronic Systems (IMS).}
\label{fig:lowTsetup}
\end{figure}
Our broad band measurement method allows to directly measure the dielectric losses by using lumped element resonators, in which the entire electric energy is stored in the material of interest between the capacitor plates with a homogeneous electric field distribution. Several LC-resonators in series are fabricated on the same chip and are measured as notch-like resonance dips (see Fig.~\ref{fig:8ressetup}).
 
This ensures that the material (and thus the mean TLS dipole moment $p$) investigated at different frequencies during one cool-down cycle is identical. In the presented measurements, a chip containing four multiplexed resonators with 400 nm of thermally evaporated a-SiO dielectric was measured. Details on the measurement method and sample fabrication are given elsewhere \cite{Kaiser}. From the measured resonance frequencies and dimensions of the parallel plate capacitors we extract the value of the relative permittivity to be $\varepsilon_{\rm r}=5.7$, suggesting that the stoichiometry of the film is close to silicon monoxide SiO.
We employed a broadband cryogenic HEMT amplifier (20 MHz - 12 GHz) \cite{Wuensch1} and a power combiner with a bandwidth ranging from DC to 12~GHz, serving here as a broadband alternative to a circulator. Its transmission properties are illustrated in Fig.~\ref{fig:8ressetup}. 
The setup includes two virtually identical low-temperature microwave measurement lines which allow for cold calibration against the path having ideal transmission. The input RF-signal coming from the network analyzer (VNA) is attenuated at room temperature, at 4.2 K, and at 100 mK. A schematic of the measurement setup is sketched in Fig.~\ref{fig:lowTsetup}.

The transmitted signal $|S_{21}|$ detected by the VNA was fitted to Lorentzian curves using a non-linear least squares fit. In the case of resonance dips, the lowest 3 dB of the dips are sufficient to be fitted to \cite{Anlage}
\begin{equation}	|S_{21}|^{2}=A\cdot\left(1+Q_{0}^{2}\left(\frac{f}{f_{0}}-\frac{f_{0}}{f}\right)^{2}\right)
  \label{eqn:Lorentzian}
\end{equation}
to yield the fitting parameter $A$ and the unloaded quality factor $Q_0$ of the resonators, with the latter being directly related to the dielectric loss in the capacitors as $\tan\delta=1/Q_{0}$ \cite{Kaiser}.
\begin{figure}
\includegraphics[width=0.85\linewidth]{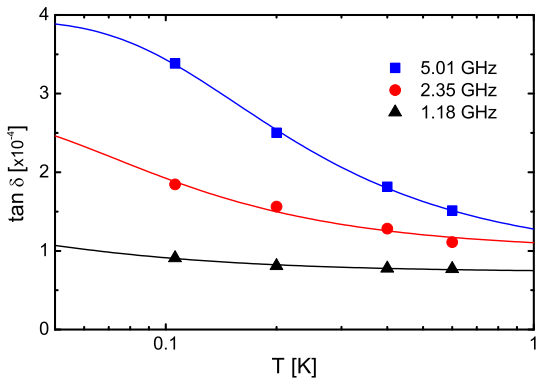}
\caption{Loss data vs. temperature of each resonator at a power level of -110 dBm. Solid lines are fits to Eq.~\eqref{eqn:tandeltastrong} with $\mathcal F(P)=1$.}
\label{fig:tandeltaFit_T}
\end{figure}
With the same technique, we have recently measured the frequency dependence of dielectric losses in various amorphous dielectrics at 4.2~K \cite{Kaiser}. At this temperature, the resonant TLS states are equally occupied and the dielectric loss is dominated by relaxation processes. This means that the electric field couples to a broad spectral distribution of TLS. In order to probe the energy dependence of the TLS density of states, we need to enter the regime dominated by resonant absorption and therefore investigate the temperature and power dependences of $\tan\delta$ as given by Eqs.~(1) and (2). First, we measured the temperature dependence of the dielectric losses at different power levels. As can be seen in Fig.~\ref{fig:tandeltaFit_T} for a power level of -110 dBm the measurement data are in agreement with fits obtained for Eq.~\eqref{eqn:tandeltastrong} in the low power limit, i.e. for $\mathcal F(P)=1$. Table \ref{table:table_temp} presents the values of $\alpha$ and $\tan\delta_{\rm sat}$ determined by fits of Eq.~\eqref{eqn:tandeltastrong} to our data at -110 dBm and -95 dBm. 
\begin{table}[h]
\caption{Values of $\tan\delta_{\rm sat}$ and $\alpha$ with determined from fits to Eq.~\eqref{eqn:tandeltastrong}.}
\begin{ruledtabular} \begin{tabular}{cccc}
Power level & $f_{0}$ [GHz] & $\tan\delta_{\rm sat}$ & $\alpha$ \\\hline
 &1.18 & $7.3\cdot 10^{-5}$ & $6.60\cdot 10^{-5}$ \\
-110 dBm &2.35 & $1.0\cdot 10^{-4}$ & $1.80\cdot 10^{-4}$ \\
$=1\cdot 10^{-14}$ W &5.01 & $9.17\cdot 10^{-5}$ & $3.02\cdot 10^{-4}$ \\ \hline
 &1.18 & $7.3\cdot 10^{-5}$ & $4.96\cdot 10^{-5}$ \\
-95 dBm &2.35 & $1.0\cdot 10^{-4}$ & $2.17\cdot 10^{-4}$ \\
$=3.16\cdot 10^{-13}$ W &5.01 & $9.17\cdot 10^{-5}$ & $2.87\cdot 10^{-4}$ \\ \end{tabular} \end{ruledtabular}
\label{table:table_temp}
\end{table}
\begin{figure}
\includegraphics[width=0.85\linewidth]{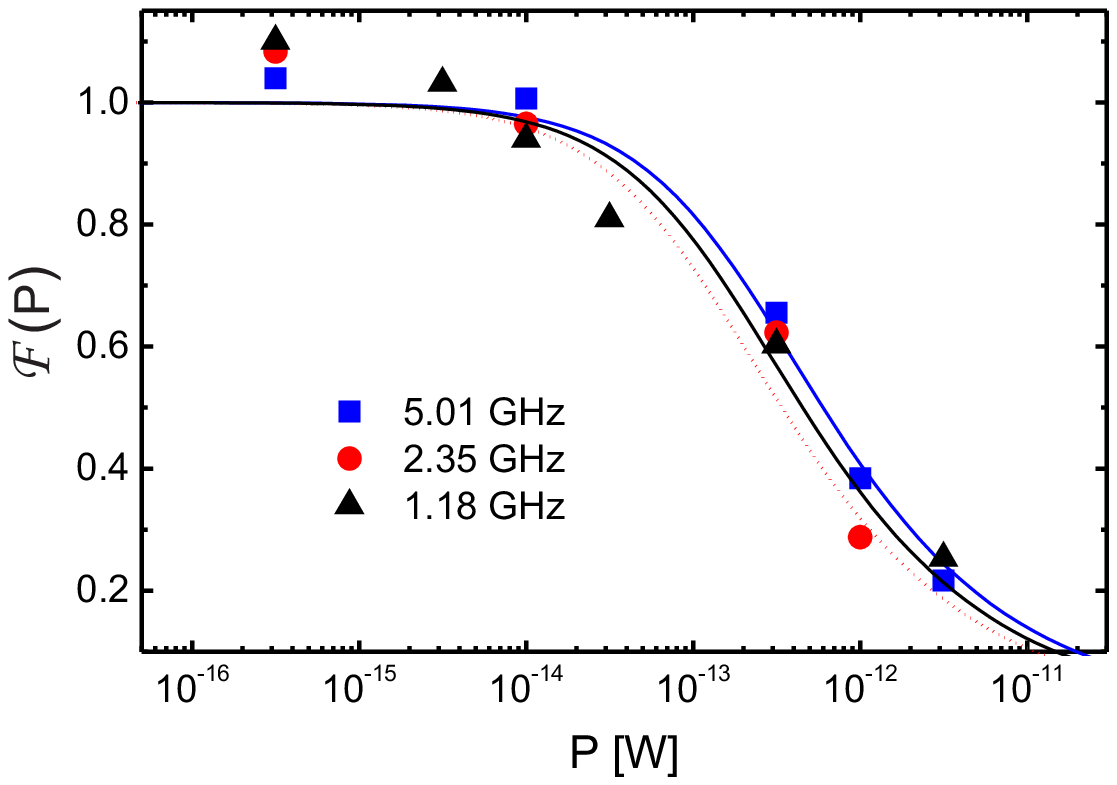}
\caption{Power dependences $\mathcal F(P)$ of dielectric losses at $T=106$ mK for all measured frequencies, with fits to Eq.~\eqref{eqn:fP}. Below $\approx10^{-14}$~W, the TLS occupation numbers remain in thermal equilibrium.}
\label{fig:tandeltaFit_P}
\end{figure}
As a next step, we investigated the power dependence $\mathcal F(P)$ of the dielectric losses. For normalization, Eq.~\eqref{eqn:tandeltastrong} was rewritten as $\mathcal F(P)=(\tan\delta-\tan\delta_{\rm sat})/\alpha\cdot\tanh(\hbar\omega/2k_{\rm B}T)$ and fitted to our measurements as shown in Fig.~\ref{fig:tandeltaFit_P}. One can see that the measured dielectric losses are in an overall good agreement with the predictions for resonant absorption by TLS. Consequently, the extracted TLS density of states $n(E)$ is expected to be proportional to the frequency dependent values $\alpha(f)$.  

\begin{figure}
\includegraphics[width=0.85\linewidth]{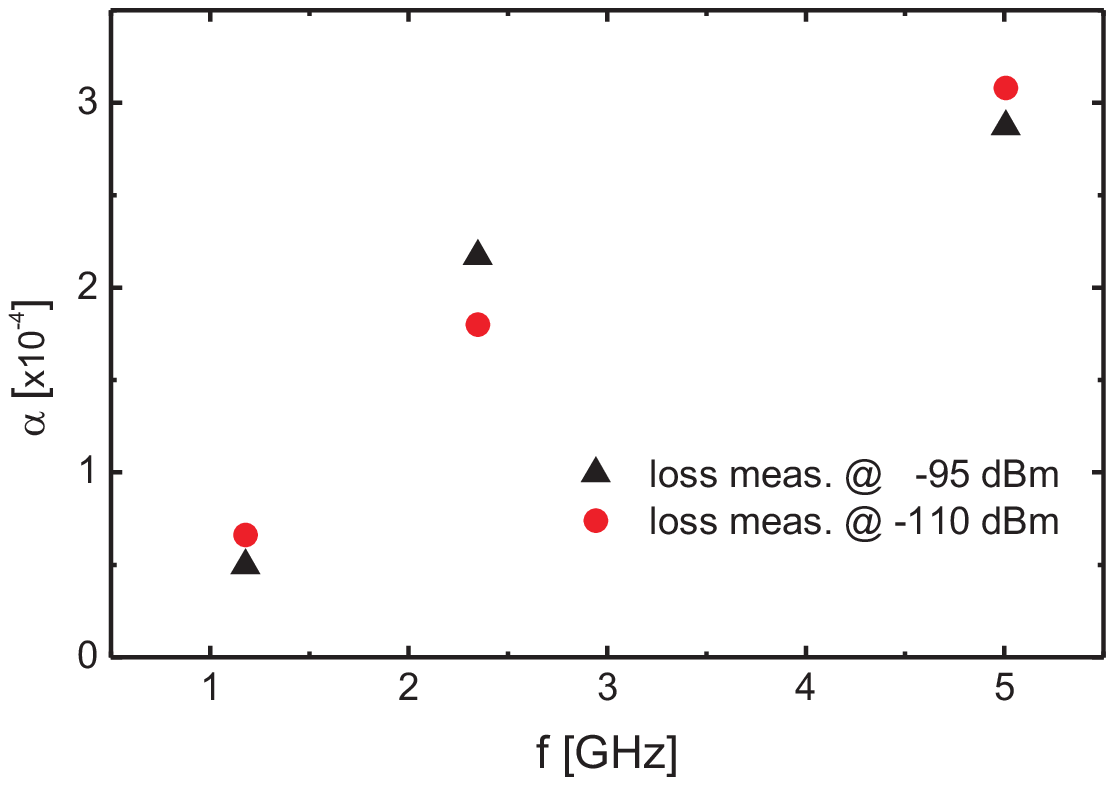}
\caption{Values of $\alpha$ obtained from fits to Eq.~\eqref{eqn:tandeltastrong} plotted over frequency $f$. The term $\alpha(f)$ is directly proportional to $n(E)$.}
\label{fig:alpha_f}
\end{figure}

As the major result of this paper, the extracted values $\alpha(f)$ are plotted in Fig.~\ref{fig:alpha_f} for all our measurements. These data suggest that the TLS density of states in our amorphous SiO films monotonically increases with energy. This result is obtained from measurements in the low power limit ($P<10^{-14}$~W) as well as at intermediate power levels ($P=3.16\cdot 10^{-13}$~W), where loss measurements are far less subject to noise and therefore more reliable. 

The extracted non-uniform density of states $n(E)$ is consistent with the following observation (not shown here): Tracking the resonance frequencies of our lumped element resonators below 1~K, we find these frequencies to decrease with decreasing temperature at gradients that become considerably smaller around 100~mK. Such a behavior is explained by the real part of $\varepsilon_{\rm r}$ of which the temperature dependence is also proportional to the TLS density of states and their changing occupation number differences \cite{Schickfus, Enss}.   

\begin{figure}[h]
\includegraphics[width=0.85\linewidth]{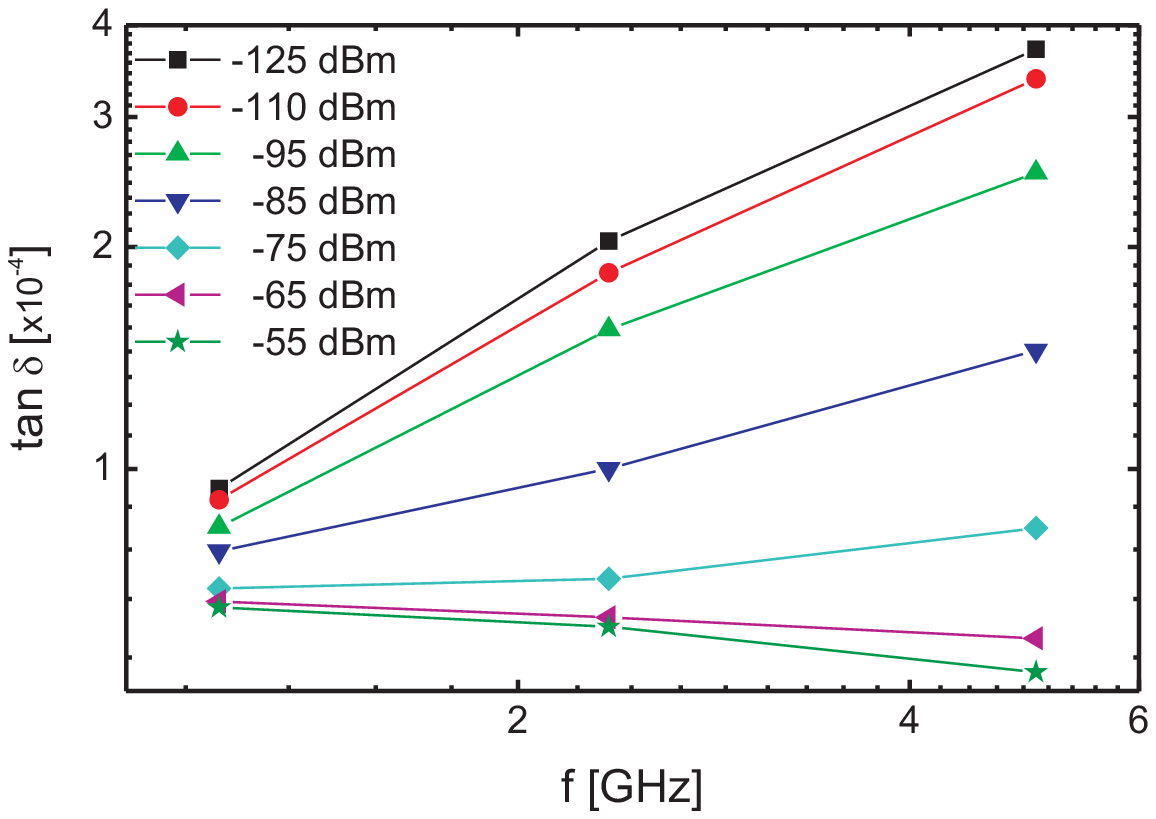}
\caption{Measured dielectric loss values at 106 mK at various power levels plotted over frequency. The low power limit where resonant absorption by TLS is dominant can be observed for $P<-110$~dBm. The high power limit which is dominated by the non-resonant part of the loss starts for $P>-65$~dBm.}
\label{fig:tandelta_f}
\end{figure}

Another result of evaluation of our experimental data is that $\tan\delta_{\rm sat}$, the non-resonant part of the dielectric loss is almost independent of frequency, as can be seen in Fig.~\ref{fig:tandelta_f} (for $P>-65$~dBm) and of temperature which can be inferred from the high temperature side in Fig.~\ref{fig:tandeltaFit_T}. In principle, we might expect both a frequency and a temperature dependence if $\tan\delta_{\rm sat}$ is at least partially caused by a relaxation absorption usually observed in glasses and well understood in terms of a non-resonant interaction of an oscillating electric field with an ensemble of TLS. Thus, our result may be taken as another hint that the TLS density of states of our a-SiO films deviates from those of glasses.\\
In conclusion, we have investigated the frequency dependence of dielectric losses of bulk a-SiO at qubit operation frequencies and temperatures where resonant absorption by TLS is dominant. Measuring frequencies are defined by superconducting lumped element resonators fabricated on a single chip, which in combination with a broadband setup allows for a frequency multiplexed measurement in a single cooling cycle.
We observed a strong energy dependence of the TLS density of states, which to our knowledge has not been reported before for loss measurements of amorphous dielectrics. These results contribute to understanding decoherence mechanisms of superconducting quantum bits.\\ \\
This work was supported in part by the EU project SOLID, the Deutsche Forschungsgemeinschaft (DFG) and the State of Baden-W\"urttemberg through the DFG Center for Functional Nanostructures (CFN). S.T.S. acknowledges support by the Heinrich B\"oll Stiftung.\\ We thank L. Faoro, T. Wirth, P. Macha, A. Bruno and M. Lisitskiy for valuable discussions as well as A. Stassen, K.-H. Gutbrod and R. Jehle for support in sample assembly, mechanical as well as electronics matters.
%


\begin{thebibliography}{19}%
\makeatletter
\providecommand \@ifxundefined [1]{%
 \@ifx{#1\undefined}
}%
\providecommand \@ifnum [1]{%
 \ifnum #1\expandafter \@firstoftwo
 \else \expandafter \@secondoftwo
 \fi
}%
\providecommand \@ifx [1]{%
 \ifx #1\expandafter \@firstoftwo
 \else \expandafter \@secondoftwo
 \fi
}%
\providecommand \natexlab [1]{#1}%
\providecommand \enquote  [1]{``#1''}%
\providecommand \bibnamefont  [1]{#1}%
\providecommand \bibfnamefont [1]{#1}%
\providecommand \citenamefont [1]{#1}%
\providecommand \href@noop [0]{\@secondoftwo}%
\providecommand \href [0]{\begingroup \@sanitize@url \@href}%
\providecommand \@href[1]{\@@startlink{#1}\@@href}%
\providecommand \@@href[1]{\endgroup#1\@@endlink}%
\providecommand \@sanitize@url [0]{\catcode `\\12\catcode `\$12\catcode
  `\&12\catcode `\#12\catcode `\^12\catcode `\_12\catcode `\%12\relax}%
\providecommand \@@startlink[1]{}%
\providecommand \@@endlink[0]{}%
\providecommand \url  [0]{\begingroup\@sanitize@url \@url }%
\providecommand \@url [1]{\endgroup\@href {#1}{\urlprefix }}%
\providecommand \urlprefix  [0]{URL }%
\providecommand \Eprint [0]{\href }%
\providecommand \doibase [0]{http://dx.doi.org/}%
\providecommand \selectlanguage [0]{\@gobble}%
\providecommand \bibinfo  [0]{\@secondoftwo}%
\providecommand \bibfield  [0]{\@secondoftwo}%
\providecommand \translation [1]{[#1]}%
\providecommand \BibitemOpen [0]{}%
\providecommand \bibitemStop [0]{}%
\providecommand \bibitemNoStop [0]{.\EOS\space}%
\providecommand \EOS [0]{\spacefactor3000\relax}%
\providecommand \BibitemShut  [1]{\csname bibitem#1\endcsname}%
\let\auto@bib@innerbib\@empty
\bibitem [{\citenamefont {Makhlin}, \citenamefont {Sch\"on},\ and\
  \citenamefont {Shnirman}(2001)}]{Schoen_review}%
  \BibitemOpen
  \bibfield  {author} {\bibinfo {author} {\bibfnamefont {Y.}~\bibnamefont
  {Makhlin}}, \bibinfo {author} {\bibfnamefont {G.}~\bibnamefont {Sch\"on}}, \
  and\ \bibinfo {author} {\bibfnamefont {A.}~\bibnamefont {Shnirman}},\
  }\href@noop {} {\bibfield  {journal} {\bibinfo  {journal} {Rev. Mod. Phys.}\
  }\textbf {\bibinfo {volume} {73}},\ \bibinfo {pages} {357} (\bibinfo {year}
  {2001})}\BibitemShut {NoStop}%
\bibitem [{\citenamefont {Martinis}\ \emph {et~al.}(2005)\citenamefont
  {Martinis}, \citenamefont {Cooper}, \citenamefont {McDermott}, \citenamefont
  {Steffen}, \citenamefont {Ansmann}, \citenamefont {Osborn}, \citenamefont
  {Cicak}, \citenamefont {Oh}, \citenamefont {Pappas}, \citenamefont
  {Simmonds},\ and\ \citenamefont {Yu}}]{Martinis}%
  \BibitemOpen
  \bibfield  {author} {\bibinfo {author} {\bibfnamefont {J.~M.}\ \bibnamefont
  {Martinis}}, \bibinfo {author} {\bibfnamefont {K.~B.}\ \bibnamefont
  {Cooper}}, \bibinfo {author} {\bibfnamefont {R.}~\bibnamefont {McDermott}},
  \bibinfo {author} {\bibfnamefont {M.}~\bibnamefont {Steffen}}, \bibinfo
  {author} {\bibfnamefont {M.}~\bibnamefont {Ansmann}}, \bibinfo {author}
  {\bibfnamefont {K.~D.}\ \bibnamefont {Osborn}}, \bibinfo {author}
  {\bibfnamefont {K.}~\bibnamefont {Cicak}}, \bibinfo {author} {\bibfnamefont
  {S.}~\bibnamefont {Oh}}, \bibinfo {author} {\bibfnamefont {D.~P.}\
  \bibnamefont {Pappas}}, \bibinfo {author} {\bibfnamefont {R.~W.}\
  \bibnamefont {Simmonds}}, \ and\ \bibinfo {author} {\bibfnamefont {C.~C.}\
  \bibnamefont {Yu}},\ }\href@noop {} {\bibfield  {journal} {\bibinfo
  {journal} {Phys. Rev. Lett.}\ }\textbf {\bibinfo {volume} {95}},\ \bibinfo
  {pages} {210503} (\bibinfo {year} {2005})}\BibitemShut {NoStop}%
\bibitem [{\citenamefont {Lindstr{\"o}m}\ \emph {et~al.}(2009)\citenamefont
  {Lindstr{\"o}m}, \citenamefont {Healey}, \citenamefont {Colclough},\ and\
  \citenamefont {Muirhead}}]{Lindstroem}%
  \BibitemOpen
  \bibfield  {author} {\bibinfo {author} {\bibfnamefont {T.}~\bibnamefont
  {Lindstr{\"o}m}}, \bibinfo {author} {\bibfnamefont {J.~E.}\ \bibnamefont
  {Healey}}, \bibinfo {author} {\bibfnamefont {M.~S.}\ \bibnamefont
  {Colclough}}, \ and\ \bibinfo {author} {\bibfnamefont {C.~M.}\ \bibnamefont
  {Muirhead}},\ }\href@noop {} {\bibfield  {journal} {\bibinfo  {journal}
  {Phys. Rev. B}\ }\textbf {\bibinfo {volume} {80}},\ \bibinfo {pages} {132501}
  (\bibinfo {year} {2009})}\BibitemShut {NoStop}%
\bibitem [{\citenamefont {Macha}\ \emph {et~al.}(2010)\citenamefont {Macha},
  \citenamefont {van~der Ploeg}, \citenamefont {Oelsner}, \citenamefont
  {Il'ichev}, \citenamefont {Meyer}, \citenamefont {W{\"u}nsch},\ and\
  \citenamefont {Siegel}}]{Macha}%
  \BibitemOpen
  \bibfield  {author} {\bibinfo {author} {\bibfnamefont {P.}~\bibnamefont
  {Macha}}, \bibinfo {author} {\bibfnamefont {S.~H.~W.}\ \bibnamefont {van~der
  Ploeg}}, \bibinfo {author} {\bibfnamefont {G.}~\bibnamefont {Oelsner}},
  \bibinfo {author} {\bibfnamefont {E.}~\bibnamefont {Il'ichev}}, \bibinfo
  {author} {\bibfnamefont {H.-G.}\ \bibnamefont {Meyer}}, \bibinfo {author}
  {\bibfnamefont {S.}~\bibnamefont {W{\"u}nsch}}, \ and\ \bibinfo {author}
  {\bibfnamefont {M.}~\bibnamefont {Siegel}},\ }\href@noop {} {\bibfield
  {journal} {\bibinfo  {journal} {Appl. Phys. Lett.}\ }\textbf {\bibinfo
  {volume} {96}},\ \bibinfo {pages} {062503} (\bibinfo {year}
  {2010})}\BibitemShut {NoStop}%
\bibitem [{\citenamefont {Gao}\ \emph {et~al.}(2008)\citenamefont {Gao},
  \citenamefont {Daal}, \citenamefont {Vayonakis}, \citenamefont {Kumar},
  \citenamefont {Zmuidzinas}, \citenamefont {Sadoulet}, \citenamefont {Mazin},
  \citenamefont {Day},\ and\ \citenamefont {Leduc}}]{Gao2}%
  \BibitemOpen
  \bibfield  {author} {\bibinfo {author} {\bibfnamefont {J.}~\bibnamefont
  {Gao}}, \bibinfo {author} {\bibfnamefont {M.}~\bibnamefont {Daal}}, \bibinfo
  {author} {\bibfnamefont {A.}~\bibnamefont {Vayonakis}}, \bibinfo {author}
  {\bibfnamefont {S.}~\bibnamefont {Kumar}}, \bibinfo {author} {\bibfnamefont
  {J.}~\bibnamefont {Zmuidzinas}}, \bibinfo {author} {\bibfnamefont
  {B.}~\bibnamefont {Sadoulet}}, \bibinfo {author} {\bibfnamefont {B.~A.}\
  \bibnamefont {Mazin}}, \bibinfo {author} {\bibfnamefont {P.~K.}\ \bibnamefont
  {Day}}, \ and\ \bibinfo {author} {\bibfnamefont {H.~G.}\ \bibnamefont
  {Leduc}},\ }\href@noop {} {\bibfield  {journal} {\bibinfo  {journal} {Appl.
  Phys. Lett.}\ }\textbf {\bibinfo {volume} {92}},\ \bibinfo {pages} {152505}
  (\bibinfo {year} {2008})}\BibitemShut {NoStop}%
\bibitem [{\citenamefont {Barends}\ \emph {et~al.}(2007)\citenamefont
  {Barends}, \citenamefont {Baselmans}, \citenamefont {Hovenier}, \citenamefont
  {Gao}, \citenamefont {Yates}, \citenamefont {Klapwijk},\ and\ \citenamefont
  {Hoevers}}]{Barends1}%
  \BibitemOpen
  \bibfield  {author} {\bibinfo {author} {\bibfnamefont {R.}~\bibnamefont
  {Barends}}, \bibinfo {author} {\bibfnamefont {J.~J.~A.}\ \bibnamefont
  {Baselmans}}, \bibinfo {author} {\bibfnamefont {J.~N.}\ \bibnamefont
  {Hovenier}}, \bibinfo {author} {\bibfnamefont {J.~R.}\ \bibnamefont {Gao}},
  \bibinfo {author} {\bibfnamefont {S.~J.~C.}\ \bibnamefont {Yates}}, \bibinfo
  {author} {\bibfnamefont {T.~M.}\ \bibnamefont {Klapwijk}}, \ and\ \bibinfo
  {author} {\bibfnamefont {H.~F.~C.}\ \bibnamefont {Hoevers}},\ }\href@noop {}
  {\bibfield  {journal} {\bibinfo  {journal} {IEEE Transaction on Applied
  Superconductivity}\ }\textbf {\bibinfo {volume} {17}},\ \bibinfo {pages}
  {263} (\bibinfo {year} {2007})}\BibitemShut {NoStop}%
\bibitem [{\citenamefont {Phillips}(1987)}]{Phillips1}%
  \BibitemOpen
  \bibfield  {author} {\bibinfo {author} {\bibfnamefont {W.~A.}\ \bibnamefont
  {Phillips}},\ }\href@noop {} {\bibfield  {journal} {\bibinfo  {journal} {Rep.
  Prog. Phys.}\ }\textbf {\bibinfo {volume} {50}},\ \bibinfo {pages} {1657}
  (\bibinfo {year} {1987})}\BibitemShut {NoStop}%
\bibitem [{\citenamefont {Phillips}(1972)}]{Phillips2}%
  \BibitemOpen
  \bibfield  {author} {\bibinfo {author} {\bibfnamefont {W.~A.}\ \bibnamefont
  {Phillips}},\ }\href@noop {} {\bibfield  {journal} {\bibinfo  {journal} {J.
  Low Temp. Phys.}\ }\textbf {\bibinfo {volume} {7}},\ \bibinfo {pages} {351}
  (\bibinfo {year} {1972})}\BibitemShut {NoStop}%
\bibitem [{\citenamefont {Anderson}, \citenamefont {Halperin},\ and\
  \citenamefont {Varma}(1972)}]{Anderson}%
  \BibitemOpen
  \bibfield  {author} {\bibinfo {author} {\bibfnamefont {P.~W.}\ \bibnamefont
  {Anderson}}, \bibinfo {author} {\bibfnamefont {B.~I.}\ \bibnamefont
  {Halperin}}, \ and\ \bibinfo {author} {\bibfnamefont {C.~M.}\ \bibnamefont
  {Varma}},\ }\href@noop {} {\bibfield  {journal} {\bibinfo  {journal} {Philos.
  Mag.}\ }\textbf {\bibinfo {volume} {25}},\ \bibinfo {pages} {1} (\bibinfo
  {year} {1972})}\BibitemShut {NoStop}%
\bibitem [{\citenamefont {Stephens}(1973)}]{Stephens}%
  \BibitemOpen
  \bibfield  {author} {\bibinfo {author} {\bibfnamefont {R.~B.}\ \bibnamefont
  {Stephens}},\ }\href@noop {} {\bibfield  {journal} {\bibinfo  {journal}
  {Phys. Rev. B}\ }\textbf {\bibinfo {volume} {8}},\ \bibinfo {pages} {2896}
  (\bibinfo {year} {1973})}\BibitemShut {NoStop}%
\bibitem [{\citenamefont {Gaganidze}\ \emph {et~al.}(1997)\citenamefont
  {Gaganidze}, \citenamefont {K{\"o}nig}, \citenamefont {Esquinazi},
  \citenamefont {Zimmer},\ and\ \citenamefont {Burin}}]{Gaganidze}%
  \BibitemOpen
  \bibfield  {author} {\bibinfo {author} {\bibfnamefont {E.}~\bibnamefont
  {Gaganidze}}, \bibinfo {author} {\bibfnamefont {R.}~\bibnamefont
  {K{\"o}nig}}, \bibinfo {author} {\bibfnamefont {P.}~\bibnamefont
  {Esquinazi}}, \bibinfo {author} {\bibfnamefont {K.}~\bibnamefont {Zimmer}}, \
  and\ \bibinfo {author} {\bibfnamefont {A.}~\bibnamefont {Burin}},\
  }\href@noop {} {\bibfield  {journal} {\bibinfo  {journal} {Phys. Rev. Lett.}\
  }\textbf {\bibinfo {volume} {79}},\ \bibinfo {pages} {5038} (\bibinfo {year}
  {1997})}\BibitemShut {NoStop}%
\bibitem [{\citenamefont {von Haumeder}, \citenamefont {Strom},\ and\
  \citenamefont {Hunklinger}(1980)}]{Haumeder}%
  \BibitemOpen
  \bibfield  {author} {\bibinfo {author} {\bibfnamefont {M.}~\bibnamefont {von
  Haumeder}}, \bibinfo {author} {\bibfnamefont {U.}~\bibnamefont {Strom}}, \
  and\ \bibinfo {author} {\bibfnamefont {S.}~\bibnamefont {Hunklinger}},\
  }\href@noop {} {\bibfield  {journal} {\bibinfo  {journal} {Phys. Rev. Lett}\
  }\textbf {\bibinfo {volume} {44}},\ \bibinfo {pages} {84} (\bibinfo {year}
  {1980})}\BibitemShut {NoStop}%
\bibitem [{\citenamefont {Kaiser}\ \emph {et~al.}(2011)\citenamefont {Kaiser},
  \citenamefont {Meckbach}, \citenamefont {Ilin}, \citenamefont {Lisenfeld},
  \citenamefont {Sch{\"a}fer}, \citenamefont {Ustinov},\ and\ \citenamefont
  {Siegel}}]{Kaiser2}%
  \BibitemOpen
  \bibfield  {author} {\bibinfo {author} {\bibfnamefont {C.}~\bibnamefont
  {Kaiser}}, \bibinfo {author} {\bibfnamefont {J.~M.}\ \bibnamefont
  {Meckbach}}, \bibinfo {author} {\bibfnamefont {K.}~\bibnamefont {Ilin}},
  \bibinfo {author} {\bibfnamefont {J.}~\bibnamefont {Lisenfeld}}, \bibinfo
  {author} {\bibfnamefont {R.}~\bibnamefont {Sch{\"a}fer}}, \bibinfo {author}
  {\bibfnamefont {A.~V.}\ \bibnamefont {Ustinov}}, \ and\ \bibinfo {author}
  {\bibfnamefont {M.}~\bibnamefont {Siegel}},\ }\href@noop {} {\bibfield
  {journal} {\bibinfo  {journal} {Supercond. Sci. Technol.}\ }\textbf {\bibinfo
  {volume} {24}},\ \bibinfo {pages} {035005} (\bibinfo {year}
  {2011})}\BibitemShut {NoStop}%
\bibitem [{\citenamefont {Zeller}\ and\ \citenamefont {Pohl}(1971)}]{Zeller}%
  \BibitemOpen
  \bibfield  {author} {\bibinfo {author} {\bibfnamefont {R.~C.}\ \bibnamefont
  {Zeller}}\ and\ \bibinfo {author} {\bibfnamefont {R.~O.}\ \bibnamefont
  {Pohl}},\ }\href@noop {} {\bibfield  {journal} {\bibinfo  {journal} {Phys.
  Rev. B}\ }\textbf {\bibinfo {volume} {4}},\ \bibinfo {pages} {2029} (\bibinfo
  {year} {1971})}\BibitemShut {NoStop}%
\bibitem [{\citenamefont {von Schickfus}\ and\ \citenamefont
  {Hunklinger}(1976)}]{Schickfus}%
  \BibitemOpen
  \bibfield  {author} {\bibinfo {author} {\bibfnamefont {M.}~\bibnamefont {von
  Schickfus}}\ and\ \bibinfo {author} {\bibfnamefont {S.}~\bibnamefont
  {Hunklinger}},\ }\href@noop {} {\bibfield  {journal} {\bibinfo  {journal} {J.
  Phys. C: Solid State Phys.}\ }\textbf {\bibinfo {volume} {9}},\ \bibinfo
  {pages} {L439} (\bibinfo {year} {1976})}\BibitemShut {NoStop}%
\bibitem [{\citenamefont {Enss}\ and\ \citenamefont {Hunklinger}(2005)}]{Enss}%
  \BibitemOpen
  \bibfield  {author} {\bibinfo {author} {\bibfnamefont {C.}~\bibnamefont
  {Enss}}\ and\ \bibinfo {author} {\bibfnamefont {S.}~\bibnamefont
  {Hunklinger}},\ }\href@noop {} {\emph {\bibinfo {title} {Low-Temperature
  Physics}}}\ (\bibinfo  {publisher} {Springer Verlag},\ \bibinfo {address}
  {Berlin Heidelberg},\ \bibinfo {year} {2005})\BibitemShut {NoStop}%
\bibitem [{\citenamefont {Kaiser}\ \emph {et~al.}(2010)\citenamefont {Kaiser},
  \citenamefont {Skacel}, \citenamefont {W{\"u}nsch}, \citenamefont {Dolata},
  \citenamefont {Mackrodt}, \citenamefont {Zorin},\ and\ \citenamefont
  {Siegel}}]{Kaiser}%
  \BibitemOpen
  \bibfield  {author} {\bibinfo {author} {\bibfnamefont {C.}~\bibnamefont
  {Kaiser}}, \bibinfo {author} {\bibfnamefont {S.~T.}\ \bibnamefont {Skacel}},
  \bibinfo {author} {\bibfnamefont {S.}~\bibnamefont {W{\"u}nsch}}, \bibinfo
  {author} {\bibfnamefont {R.}~\bibnamefont {Dolata}}, \bibinfo {author}
  {\bibfnamefont {B.}~\bibnamefont {Mackrodt}}, \bibinfo {author}
  {\bibfnamefont {A.}~\bibnamefont {Zorin}}, \ and\ \bibinfo {author}
  {\bibfnamefont {M.}~\bibnamefont {Siegel}},\ }\href@noop {} {\bibfield
  {journal} {\bibinfo  {journal} {Supercond. Sci. Technol.}\ }\textbf {\bibinfo
  {volume} {23}},\ \bibinfo {pages} {075008} (\bibinfo {year}
  {2010})}\BibitemShut {NoStop}%
\bibitem [{\citenamefont {W{\"u}nsch}\ \emph {et~al.}(2009)\citenamefont
  {W{\"u}nsch}, \citenamefont {Ortlepp}, \citenamefont {Crocoll}, \citenamefont
  {Uhlmann},\ and\ \citenamefont {Siegel}}]{Wuensch1}%
  \BibitemOpen
  \bibfield  {author} {\bibinfo {author} {\bibfnamefont {S.}~\bibnamefont
  {W{\"u}nsch}}, \bibinfo {author} {\bibfnamefont {T.}~\bibnamefont {Ortlepp}},
  \bibinfo {author} {\bibfnamefont {E.}~\bibnamefont {Crocoll}}, \bibinfo
  {author} {\bibfnamefont {F.~H.}\ \bibnamefont {Uhlmann}}, \ and\ \bibinfo
  {author} {\bibfnamefont {M.}~\bibnamefont {Siegel}},\ }\href@noop {}
  {\bibfield  {journal} {\bibinfo  {journal} {IEEE Transaction on Applied
  Superconductivity}\ }\textbf {\bibinfo {volume} {19}},\ \bibinfo {pages}
  {574} (\bibinfo {year} {2009})}\BibitemShut {NoStop}%
\bibitem [{\citenamefont {Petersan}\ and\ \citenamefont
  {Anlage}(1998)}]{Anlage}%
  \BibitemOpen
  \bibfield  {author} {\bibinfo {author} {\bibfnamefont {P.~J.}\ \bibnamefont
  {Petersan}}\ and\ \bibinfo {author} {\bibfnamefont {S.~M.}\ \bibnamefont
  {Anlage}},\ }\href@noop {} {\bibfield  {journal} {\bibinfo  {journal} {J.
  Appl. Phys.}\ }\textbf {\bibinfo {volume} {84}},\ \bibinfo {pages} {3392}
  (\bibinfo {year} {1998})}\BibitemShut {NoStop}%
\end{thebibliography}

%

\end{document}